\documentclass[11pt]{article} \textwidth 170mm \textheight 225mm
\usepackage{amsfonts}
\usepackage{amssymb}
\usepackage{dcolumn}
\usepackage{amsthm}
\usepackage{amsmath}
\usepackage{multicol}
\usepackage{newlfont}
\usepackage{newlfont}
\usepackage{multicol}
\usepackage{bm}
\usepackage{txfonts}
\usepackage{graphics}
\usepackage{graphicx}
\usepackage{color}
\usepackage{indentfirst}
\begin{document}

\title{First-principles calculations of phonon and thermodynamic properties of AlRE (RE= Y, Gd, Pr, Yb) intermetallic compounds \footnote{The work is supported by the National Natural Science Foundation
of China (11074313) and and Project No.CDJXS11102211 supported by the Fundamental Research Funds for the Central Universities of China. }}
\author{Rui Wang\footnote{Tel: +8613527528737; E-mail: rcwang@cqu.edu.cn.}, Shaofeng Wang, and Xiaozhi Wu\\
{\small  {Department of Physics and Institute for Structure and Function, Chongqing University, }}\\ {\small {
Chongqing 400044, P. R. China. }} }

\date{}

\maketitle
\begin{abstract}
\baselineskip 18pt
\noindent The phonon and thermodynamic properties of rare-earth-aluminum intermetallics AlRE (RE=Y, Gd, Pr, Yb) with B2-type structure are investigated by performing density functional theory and density functional perturbation theory within the quasiharmonic approximation. The phonon spectra and phonon density of states, including the phonon partial density of states and total density of states, have been discussed. Our results demonstrate that the density of states is mostly composed of Al states at the high frequency. The temperature dependence of various quantities such as the thermal expansions, the heat capacities at constant volume and constant pressure, the isothermal bulk modulus, and the entropy are obtained. The electronic contribution to the specific heat is discussed, and the presented results show that the thermal electronic excitation affecting the thermal properties is inessential.
\end{abstract}

\vskip 0.1in{\small   } \vskip 0.2in

\vskip 0.1in

 \vskip 0.3in

\baselineskip 24pt


\section{Introduction}

Aluminum alloys are widely used for aerospace industries, aircraft automotive industries, electronic industries and buildings due to light weight, good corrosion resistance, reasonably high strength and favorable economics. Addition of rare earth (RE) elements to Al-based metal alloys have received great attention since it can result in increasing their wear resistance \cite{Shi2010}, mechanical properties \cite{Knipling},
electrochemical behavior \cite{Rosalbinoa} and thermal stability \cite{Nie2002},etc. Thus, they are important for design of novel materials and further scientific and technical investigations. The development and design of new Aluminum alloys need more fundamental physical data of these alloys. Recently, some investigations focusing on AlRE, Al$_{2}$RE, and Al$_{3}$RE systems intensely are performed from first-principle calculations based on density functional theory (DFT) \cite{Srivastava,Tao1,Tao,Asta,Ugur,Pang2011,Zhou2010,Wang2011s}.  The B2-type AlY is a high-temperature stable while the temperature higher than 1473K \cite{Dagerhamn}. Such metastale phases are also available for AlGd, AlPr, and AlYb, etc. intermetallics with B2-type structure \cite{Villars,Baenziger,Kripjakevich}. Srivastava et. al. \cite{Srivastava}  investigated the electronic and thermal properties by using the tight binding linear muffin tin orbital (TB-LMTO). Tao et. al. \cite{Tao1,Tao} performed the projector augmented wave (PAW) method to calculate the elastic properties and predicted thermodynamical properties of B2-AlRE intermetallics below $400K$ within the Debye-type model. However, the accurate calculations of thermodynamic properties for B2-type AlRE intermetallics from the lattice dynamics, i.e. exact phonon spectra, have received little attention.

Thermodynamical properties, such as entropy, specific heat, thermal expansion, and temperature dependent equation of state (EOS) of AlRE intermetallics are very important to  investigate their properties  as well as applications. A simplified method for thermal expansion calculations is in the framework of density functional theory (DFT) with a Debye-Gr\"{u}neisen based model \cite{Moruzzi}, which is based on the long-wave approximation. This model is useful for calculating the thermal properties at low temperature, and thermodynamic properties of B2-type AlRE intermetallics have been carried out based on this model by Tao et. al \cite{Tao}. A more accurate approach has been made possible by the achievements of density functional perturbation theory (DFPT) from first principles\cite{Giannozzi}, which allowed exact calculations of vibrational frequencies in every point of the Brillouin Zone \cite{Nie}. The vibrational free energy can be obtained using the quasiharmonic approximation (QHA). Moreover, QHA method lets one take into account the anharmonicity of the potential at the first order: vibrational properties can be understood in terms of the excitation of the noninteracting phonon. QHA based on DFPT provided a reasonable description of the thermodynamical properties of many bulk materials below the melting point \cite{Nie,Biernacki,Pavone,Carrier,Togo2010}, and had been applied with great success to more and more complex materials such as alloys [NiAl$_{3}$ \cite{Wang2004}], perovskite [MgSiO$_{3}$ \cite{Karki}], and hexaborides [LaB$_{6}$  and CeB$_{6}$ \cite{Gurel}]. More recently, we have successfully investigated the thermodynamical properties of NiAl, YAg, and YCu \cite{Rui2011i}, and MgRE intermetallics \cite{Rui2011} with B2-type structures.

In this paper, we apply first principles calculations within QHA based on DFPT to study the thermodynamic properties of AlRE (RE=Y, Gd, Pr, Yb) intermetallics with B2-type structures. The phonon spectra and phonon density of states, including the phonon partial density of states and total density of states, have been discussed. Thermal expansions, temperature dependence of isothermal bulk modules, heat capacities at constant volume and constant pressure, and the entropy as a function of temperature are presented.

\section{Theory}

To study the effects of changing temperature, one has to
look at the Helmholtz free energy, incorporating the effects of thermal-electronic excitations and thermal-vibrations (phonons). The Helmholtz free energy at temperature $T$ and constant volume $V$ is give by \cite{Moriarty, Tanju}
\begin{equation}\label{Ftotal}
F(V, T)=E_{\mathrm{0}}(V)+F_{\mathrm{el}}(V,T)+F_{\mathrm{vib}}(V, T),
\end{equation}
where $E_{\mathrm{0}}(V)$ is the static contribution to the internal energy at volume V and can be easily obtained form standard DFT calculations. $F_{\mathrm{el}}(V,T)$ given in Eq. (\ref{Ftotal}) is the thermal electronic contribution to free energy and is given by $F_{\mathrm{el}}=E_{\mathrm{el}}-TS_{\mathrm{el}}$. Here,  The electronic excitation energy $E_{\mathrm{el}}$ is given by
\begin{equation}\label{ele-Energy}
E_{\mathrm{el}}(V,T)=\int_{0}^{\infty}n(\varepsilon,V)f(\varepsilon)\varepsilon d\varepsilon-\int_{0}^{\varepsilon_{F}}n(\varepsilon,V)\varepsilon d\varepsilon,
\end{equation}
where $n(\varepsilon,V)$ is the electronic density of state (DOS) at energy $\varepsilon$ and volume $V$, $f(\varepsilon)$ is the Fermi-Dirac distribution function, and $\varepsilon_{F}$ represents the Fermi level. The electronic entropy $S_{\mathrm{el}}$ is formulated as
\begin{equation}\label{ele-entropy}
S_{\mathrm{el}}(V,T)=-k_{\mathrm{B}}\int_{0}^{\infty}n(\varepsilon,V)[f(\varepsilon)\ln f(\varepsilon) +(1-f(\varepsilon))\ln (1-f(\varepsilon))] d\varepsilon,
\end{equation}
where $k_{\mathrm{B}}$ is the Boltzmann constant. Thermal electronic contribution to free energy is generally considered to be negligible away from the melting point of the material under consideration.

$F_{\mathrm{vib}}(V, T)$ Eq. (\ref{Ftotal}) is the vibrational free energy which comes from the phonon contribution. Within the quasiharmonic approximation (QHA), $F_{\mathrm{vib}}(V, T)$ is given by
\begin{equation}\label{Fph}
F_{{\mathrm{vib}}}(V, T)=k_{B}T\sum_{\mathbf{q},\lambda}\ln\Bigg\{2\sinh\bigg(\frac{\hbar\omega_{\mathbf{q},\lambda}(V)}{2k_{B}T}\bigg)\Bigg\}.
\end{equation}
Here, the sum is over all phonon branches $\lambda$ and over all wave vectors $\mathbf{q}$ in the first Brillouin zone, $\hbar$ is the reduced Planck constant, and $\omega_{\mathbf{q}\lambda}(V)$ is the frequency of the phonon with wave vector $\mathbf{q}$ and polarization $\lambda$, evaluated at constant volume $V$.

The vibrational specific heat $C_{\mathrm{V}}$ at constant volume in the QHA from the following equation
\begin{equation}\label{cvib}
C_{\mathrm{V}}^{\mathrm{vib}}=\sum_{\mathbf{q}\lambda}k_{\mathrm{B}}\bigg(\frac{\hbar\omega_{\mathbf{q}\lambda}(V)}{2k_{\mathrm{B}}T}\bigg)^2 {\cosh^{2}\bigg(\frac{\hbar\omega_{\mathbf{q}\lambda}(V)}{k_{\mathrm{B}}T}\bigg)^{2}}.
\end{equation}
The electronic specific heat can be obtained from
\begin{equation}\label{cel}
C_{\mathrm{V}}^{\mathrm{el}}=T\bigg(\frac{\partial S_{\mathrm{el}}}{\partial T}\bigg)_{V},
\end{equation}
and we denote that total specific heat at constant volume is then $C_{\mathrm{V}}=C_{\mathrm{V}}^{\mathrm{ph}}+C_{\mathrm{V}}^{\mathrm{el}}$.
Due to anharmonicity, the specific heat at a constant pressure, $C_{p}$, is different from the specific heat at a constant volume, $C_{V}$ goes to a constant which is given by classical equipartition law: $C_{V}=3Nk_{B}$, where $N$ is the number of atoms in the system while $C_{p}$, which is what experiments determine directly, is proportional to $T$. QHA lets one take into account the anharmonicity of the potential at first order: vibrational
properties can be understood in terms of the excitation of the noninteracting phonon. The equilibrium volume at temperature $T$ is obtained by minimizing Helmholtz $F$ with respect to $V$, i.e., $\min_{V}[F(V,T)]$. The volume thermal expansion coefficient is given by
\begin{equation}
\alpha(T)=\frac{1}{V}\bigg(\frac{\partial V}{\partial T}\bigg)_{P},
\end{equation}
and the linear thermal expansion is described as
\begin{equation} \label{epsilon}
\epsilon(T)=\frac{a(T)-a(T_{c})}{a(T_{c})},
\end{equation}
where $a(T_{c})$ is equilibrium lattice constant $a(T)=[V(T)]^{1/3}$  at $T_{c}=300K$.

Then, $C_{p}$ can be obtained from $C_{V}$ and $\alpha$ by
\begin{equation}\label{cp}
C_{p}=C_{V}+\alpha^{2}BVT,
\end{equation}
where $B(T)=-1/V\partial^2 F/\partial V^2$ is the bulk modulus.

The vibrational contribution to the entropy of the crystal is given by
\begin{equation}
S_{\mathrm{vib}}=-k_{\mathrm{B}}\sum_{\mathbf{q}\lambda}\Bigg[\ln \bigg(2\sinh\frac{\hbar\omega_{\mathbf{q}\lambda}(V)}{2k_{\mathrm{B}}T}\bigg)-
\frac{\hbar\omega_{\mathbf{q}\lambda}(V)}{2k_{\mathrm{B}}T}\coth\frac{\hbar\omega_{\mathbf{q}\lambda}(V)}{2k_{\mathrm{B}}T}\Bigg].
\end{equation}

\section{Computational details}

In present work, the static energy and the thermal electronic contribution to the Helmhotz free energy are computed
by using the first-principles calculations in the framework of the density-functional theory (DFT).
We employed the generalized gradient approximation (GGA) in the
Perdew-Burke-Ernzerhof (PBE) \cite{Perdew1,Perdew2} exchange-correlation functional as implemented in the VASP code \cite{Kresse1, Kresse2,
Kresse3}. The ion-electron interaction is described by the full potential frozen-core projector augmented wave (PAW) method \cite{Blochl, Kresse4}, with energy cutoff of 600eV for plane waves. The Brillouin zones of the unit cells are represented by
Monkhorst-Pack special k-point scheme \cite{Monkhorst}. We used a sample as $21\times21\times21$ $k$-point mesh in the full Brillouin zone giving 726 irreducible $k$-points to calculate the initial structures and electronic DOS. The radial cutoffs of the PAW potentials of Al, Y, Gd, Pr and Yb were 1.40, 1.81, 1.58, 1.64, and 1.74 {\AA}, respectively. The 3$s$ and 3$p$ electrons for Al, the 4$s$, 4$p$, 4$d$ and 5$s$ electrons for Y, the 4$f$ and 6$s$ electrons for Gd, Pr, and Yb were treated as valence and the remaining electrons were kept frozen. The thermal electronic energies and entropies are evaluated using one-dimensional integrations from the self-consistent DFT calculations of electronic DOS using Fermi-Dirac smearing as shown in Eqs. (\ref{ele-Energy}) and (\ref{ele-entropy}). In order to deal with the possible convergence problems for metals, a smearing technique is employed using the Methfessel-Paxton scheme \cite{Methfessel}, with a smearing with of 0.05eV.

 The vibrational free energy were obtained from the first-principles phonon calculations by using PHONOPY \cite{Togo2010,Togo2008,Togop} which can support VASP interface to calculate force constants directly in the framework of density-functional perturbation theory (DFPT) \cite{Kresse5}. Phonon calculations were performed by the supercell approach. Since the chosen supercell size strongly influences on the thermal properties, we compare the vibrational free energies of $3\times3\times3$ supercell with those of $5\times5\times5$ supercell at 300K and 1000K, and find that the energy fluctuations between $3\times3\times3$ and $5\times5\times5$ supercells are less than 0.01\%. Hence, we chose the $3\times3\times3$ supercell with 54 atoms to calculate phonon dispersions. We carried out DFPT calculations on this 54 atoms supercell using PBE-GGA exchange-correlations effects and $7\times7\times7$ $k$-point grid meshes for Brillouin zone integrations.

\begin{figure}
\scalebox{0.7}[0.7]{\includegraphics{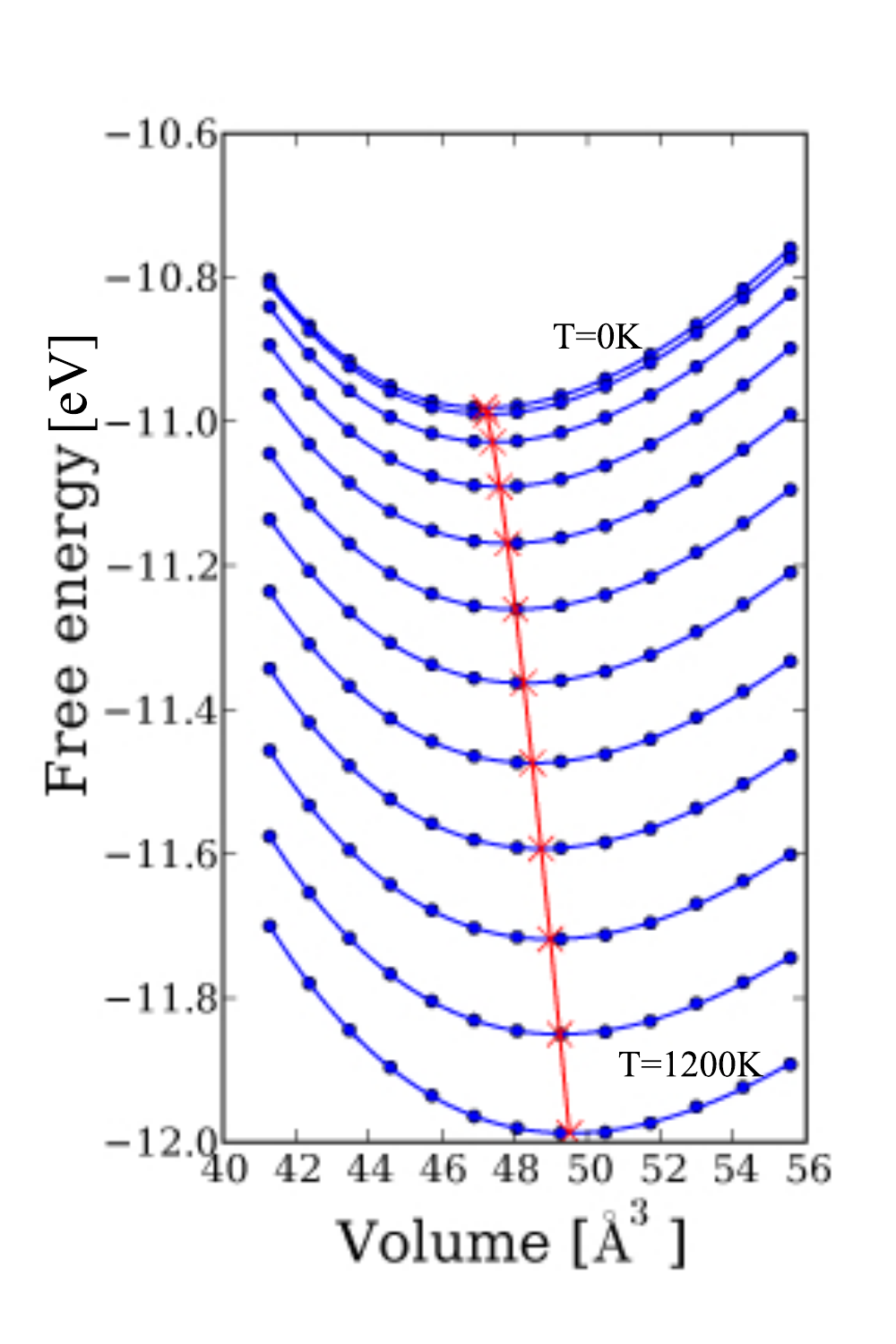}}
\caption{(Color online) The Helmholtz free energy $F(V, T)$ as a function of unit-cell volume of AlY. The circles denote $F(V, T)$ at the volume points at every 100K between 0 and 1200K. The solid curves show the fitted thermodynamic functions from Vinet EOS. The minimum values of the fitted thermodynamic functions at temperatures are depicted by the crosses. The dashed curve passing through the crosses is guide to the eye.   }
\label{EOS}
\end{figure}

 In order to get the equilibrium lattice volume as a function of temperature, we have calculated total free energy at temperature points with a step of 1K from 0 to 1200K at 13 volume points. At each temperature point, the equilibrium volume $V(T)$ and isothermal bulk moduli B(T) are obtained by minimizing free-energy with respect to $V$ from fitting the integral form of the Vinet equation of state (EOS) \cite{Vinet} at $p=0$. These procedures applied for AlY are demonstrated in Figure \ref{EOS}, where the Helmholtz free energy $F(T, V)$ as a function of unit-cell volume at temperatures are shown. While, $C_{p}$ has be calculated  by polynomial fittings for $C_{\mathrm{V}}$ and by numerical differentiation for $\partial V / \partial T$ to obtain $\alpha (T)$.

\section{Results and discussions}
\subsection {Phonon spectra and density of states}
\begin{figure}

\scalebox{0.35}[0.35]{\includegraphics{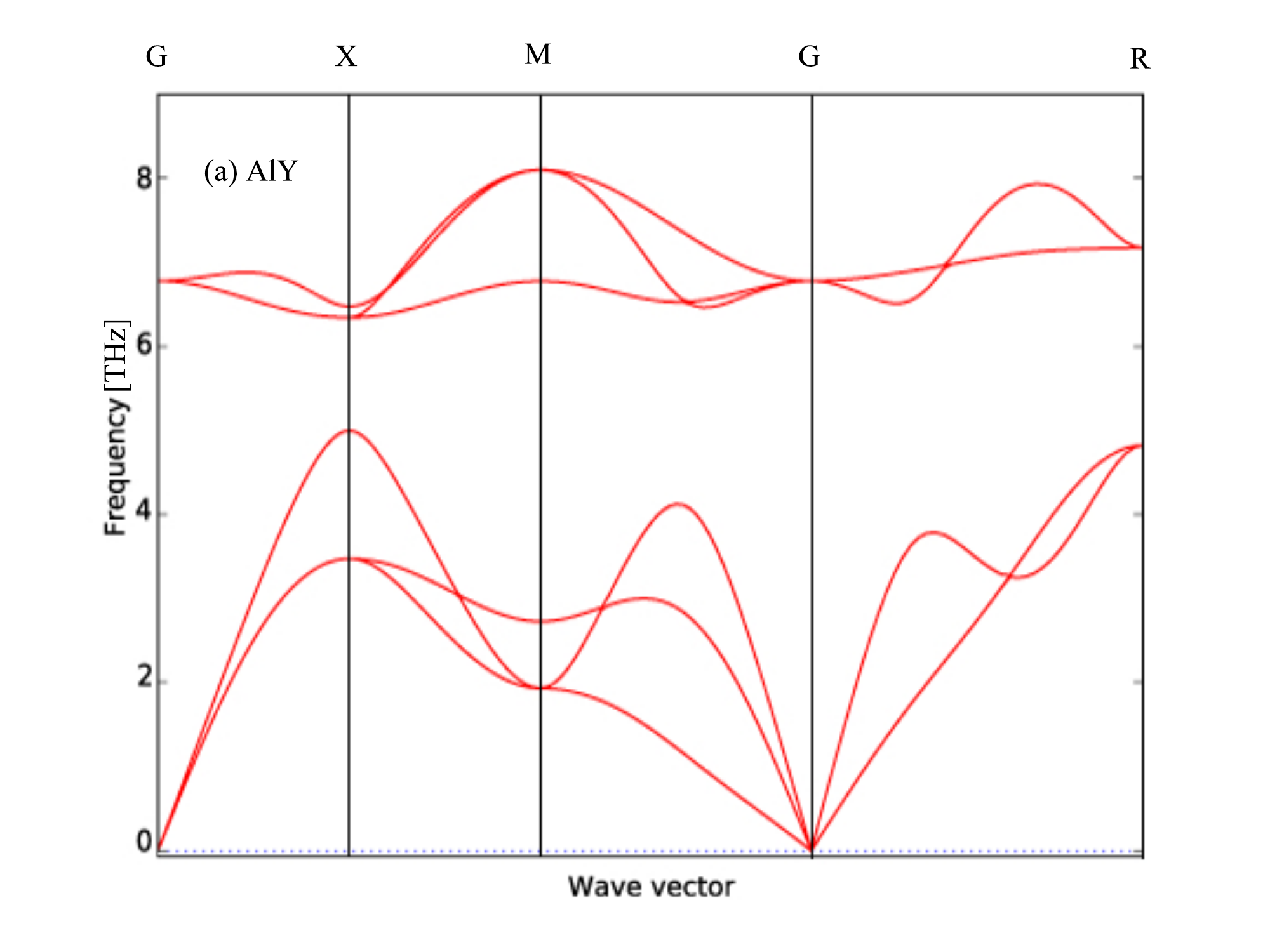}}
\scalebox{0.35}[0.35]{\includegraphics{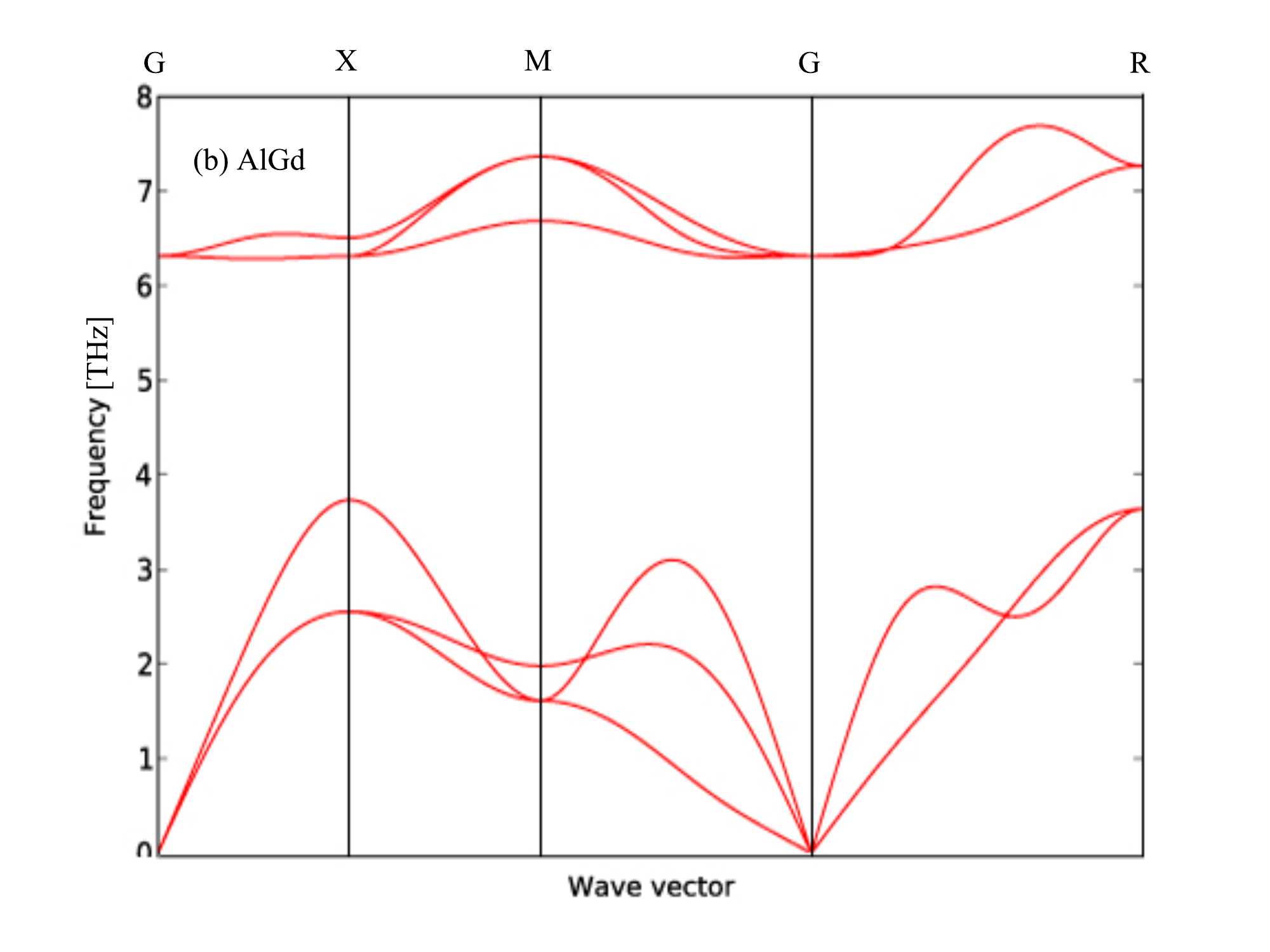}}

\scalebox{0.35}[0.35]{\includegraphics{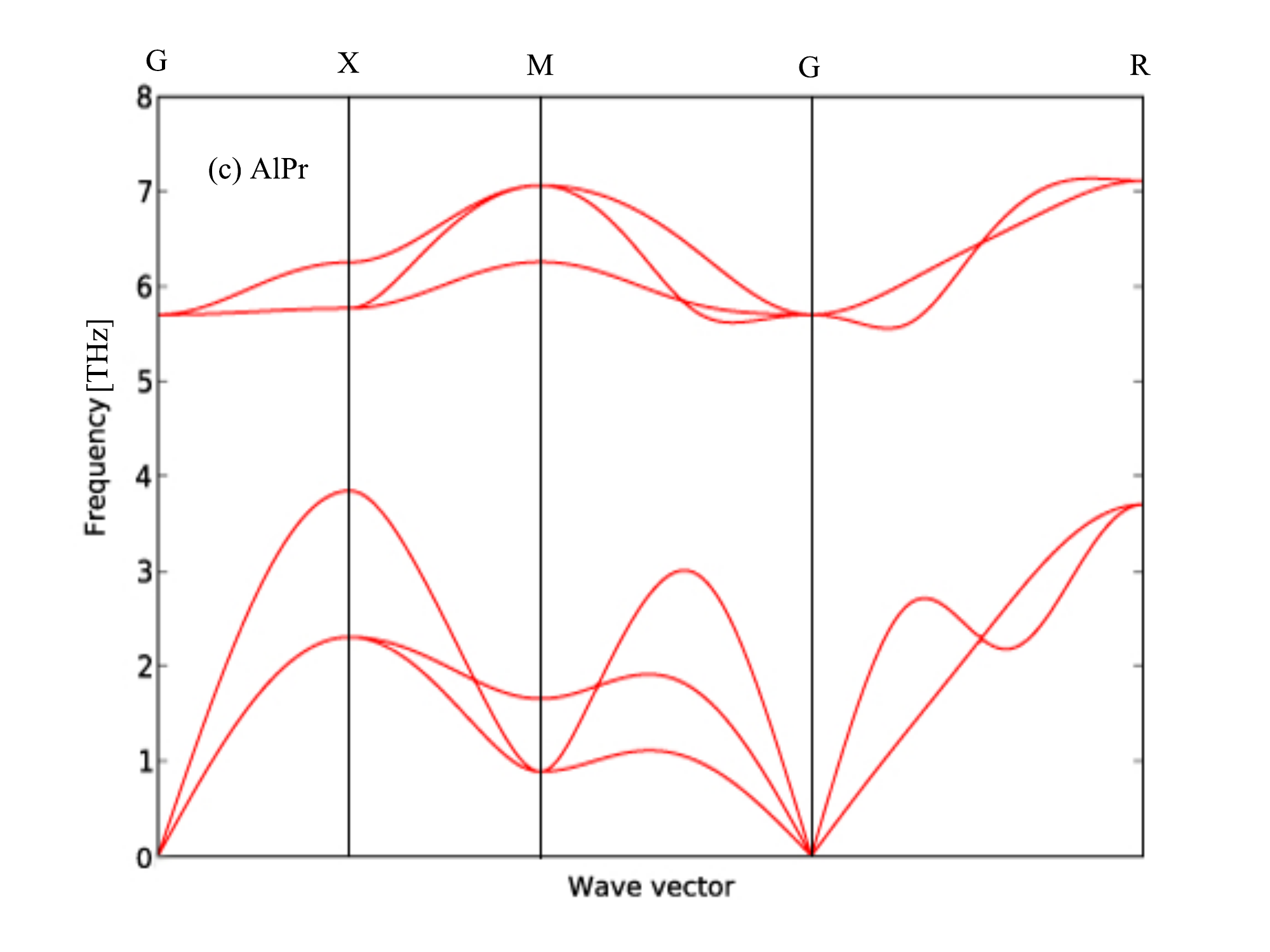}}
\scalebox{0.35}[0.35]{\includegraphics{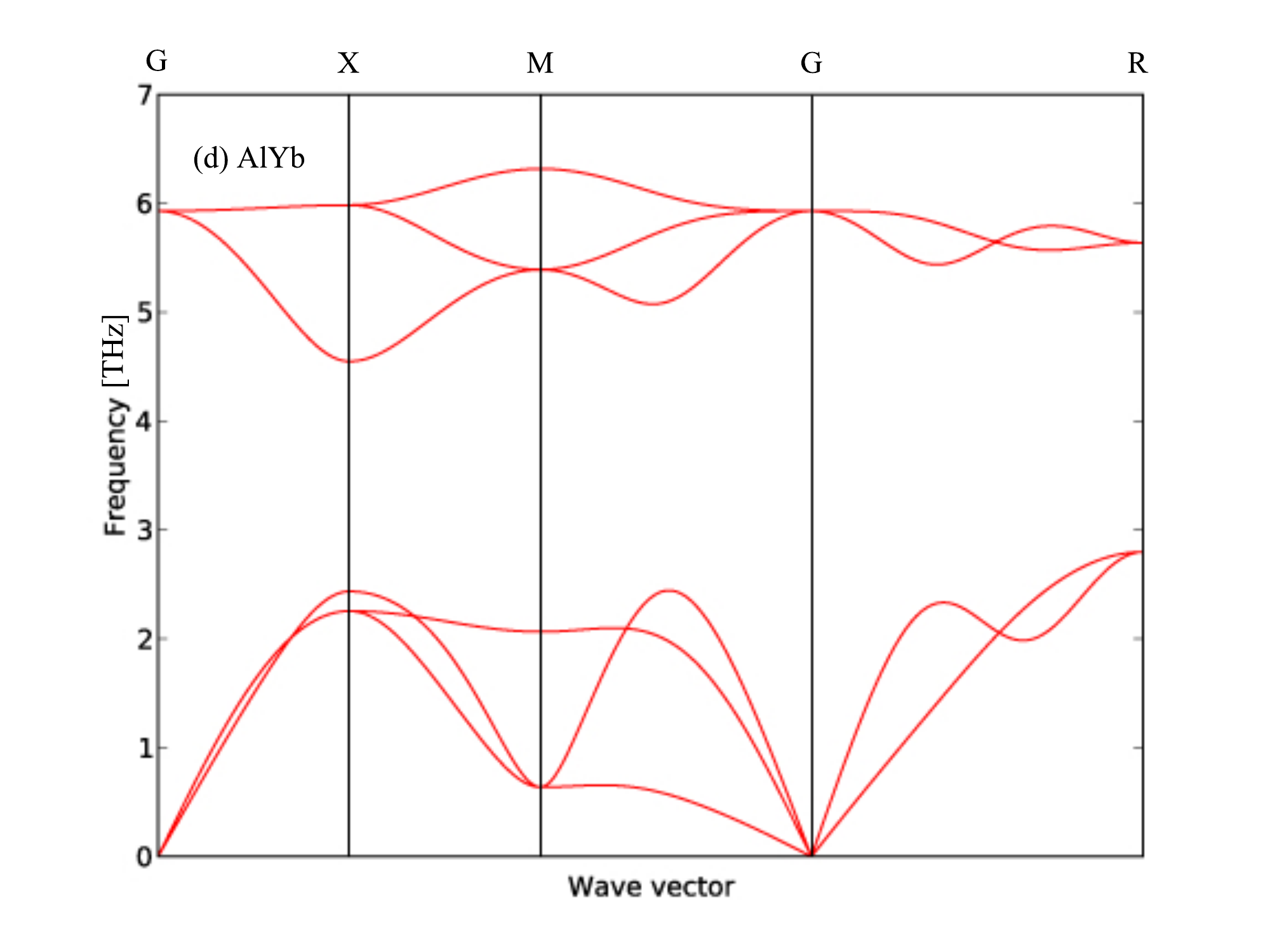}}
\caption{(Color online) Phonon-dispersion curves of (a) AlY, (b) AlGd, (c) AlPr, and (d) AlYb. }
\label{phonon}
\end{figure}

\begin{figure}
\scalebox{0.7}[0.7]{\includegraphics{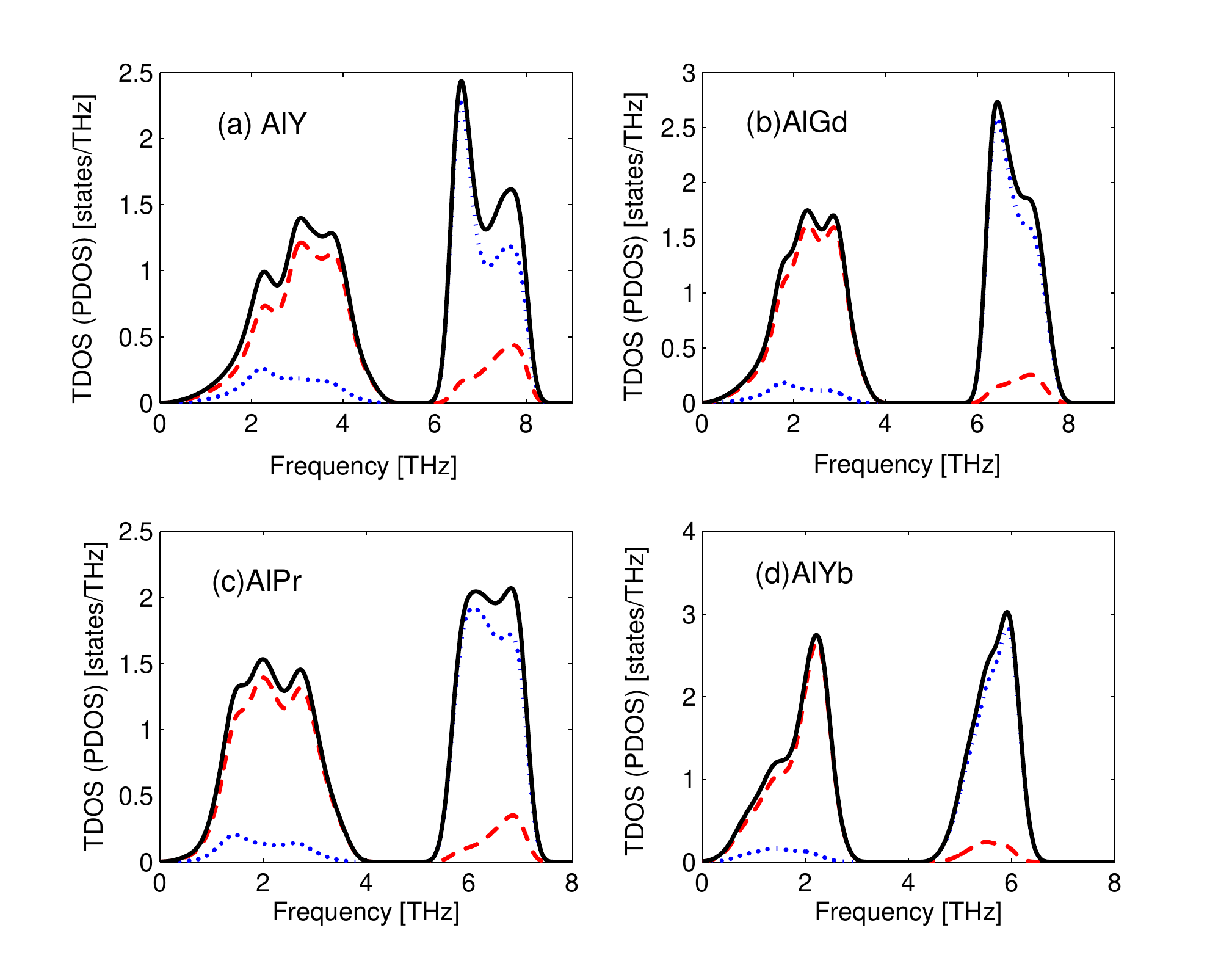}}
\caption{Phonon TDOS and PDOS of (a) AlY, (b) AlGd, (c) AlPr, and (d) AlYb. The solid curves, dashed curves, and dotted curves represent the phonon total DOS (TDOS), partial DOS (PDOS) of RE (RE=Y, Gd, Pr, and Yb) states, and PDOS of Al states, respectively. The PDOS indicates that the DOS are mostly composed of Al states at high frequency and RE states at low frequency.}
\label{dos}
\end{figure}

We calculated phonon frequency using the DFPT with force-constant method \cite{Alfe} with forces calculated using VASP.  The calculated phonon spectra of AlY, AlGd, AlPr, and AlYb, which are calculated by using $3\times3\times3$ supercells, are displayed in Figures \ref{phonon}. Due to the crystal symmetry, the spectra curves are shown along high-symmetry direction $G-X-M-\Gamma-R$ of the Brillouin zone. These dispersion curves have common framework. Since the B2-type intermetallics contain two atoms per primitive cubic unit cell, there are three acoustical branches and three optical branches. Our results obviously show that the degenerating of two transverse-acoustical (TA) modes and two transverse-optical (TO) modes in the $G-X$ and $G-R$ directions. At the $G$ point, the degenerating of longitudinal-optical (LO) and TO modes is obtained, so there are no polarization effects in the B2-type AlRE intermetallics. The phonon DOS including the partial DOS (PDOS) and the total DOS (TDOS) are shown in Figures \ref{dos}.  Sampling a $51\times51\times51$ Monkhorst-Pack grid for phonon wave vectors $\mathbf{q}$ is found to be sufficient in order to get the mean relative error in each channel of phonon DOS.  The phonon band gap of AlY starts at 5 THz, while those of AlGd, AlPr, and AlYb starts at less than 4Hz. Among the four intermetallics, AlY and AlYb has the greatest and lowest values of the maximum value of acoustic modes, respectively, since the atomic mass of Y is lightest and that of Yb is heaviest in the four calculated RE elements. The flat regions of phonon-dispersion curves, which correspond to the peaks in the phonon PDOS, indicate localization of  the states, i.e., they behave like "atomic states" \cite{Togo2010}. In the four intermetallics, the density of states are mostly composed of Al states above the phonon band gap since its atomic mass is much lighter than those of the rare earth elements RE (RE=Y, Dy, Pr, Tb).

\subsection {Bulk properties and thermal expansion}
\begin{figure}
\scalebox{0.7}[0.7]{\includegraphics{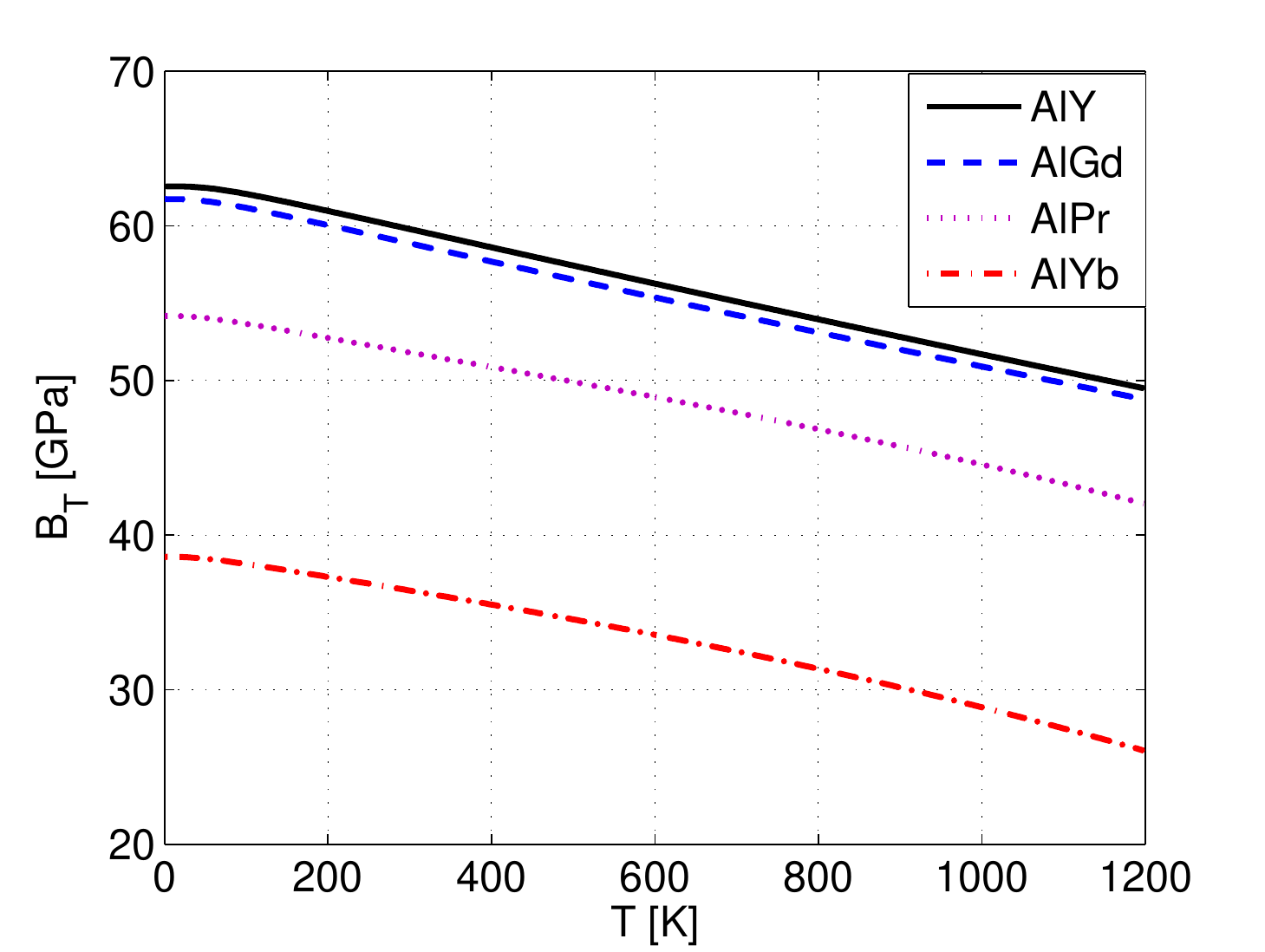}}
\caption{(Color online) Isothermal bulk moduli $B$ as a function of temperature. }
\label{Bt}
\end{figure}

The calculated results of the equilibrium lattice constants $a_{0}$, the isothermal bulk modulus $B_{0}$, and the pressure derivatives of the isothermal bulk modulus $B_{0}'$ at $T=0K$ for AlRE (RE=Y, Dy, Pr, Tb) intermetallics together with  the previous calculated results \cite{Tao1} and the available experimental values \cite{Dagerhamn,Baenziger,Kripjakevich} are shown in Table \ref{table}. Our calculated results for the equilibrium lattice constants at $T=0K$  show excellent agreements with the previous theoretical and experimental results. For the isothermal bulk modulus $B_{0}$ and the pressure derivatives of the isothermal bulk modulus $B_{0}'$ at $T=0$, which are obtained from Vinet equation of state, the present results are within 1.7\% errors from the previous calculated values obtained from the Rose's equation of state \cite{Tao1}. The temperature dependent isothermal bulk modulus $B(T)$ are shown in Figure \ref{Bt}. Among the four intermetallic compounds, through the temperature range $0-1200K$, lihgt RE AlY and heavy RE AlYb have the highest and lowest bulk moduli, respectively, i.e, AlY is the most incompressible and AlYb is the most compressible. With increasing temperature, the differences among the bulk moduli almost remain unchanged. The overall observation is that  $B_{T}$ of the intermetallics decrease with increasing temperature, and approach linearity at higher temperature and zero slope around zero temperature.

\begin{figure}
\scalebox{0.5}[0.5]{\includegraphics{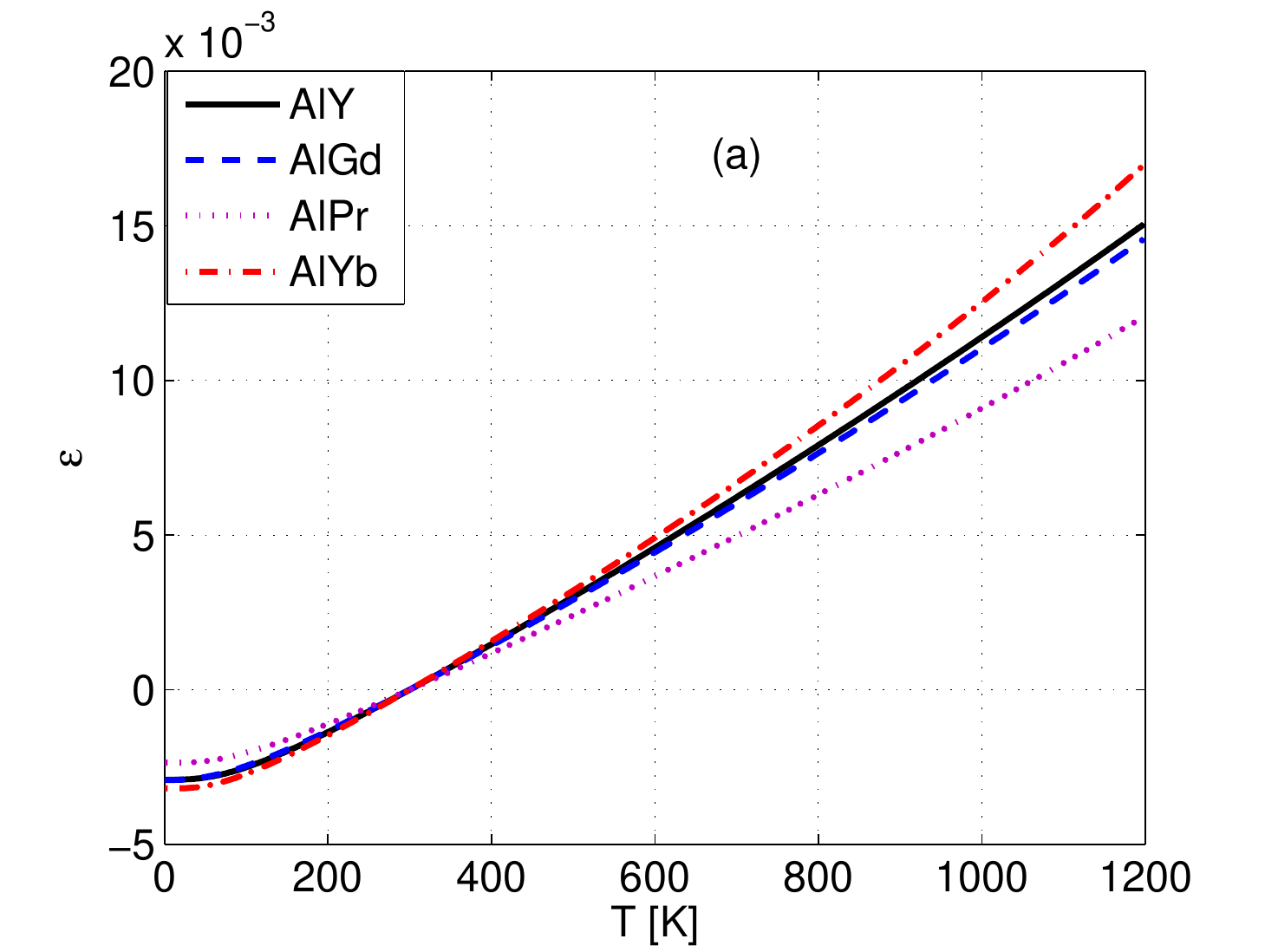}}
\scalebox{0.5}[0.5]{\includegraphics{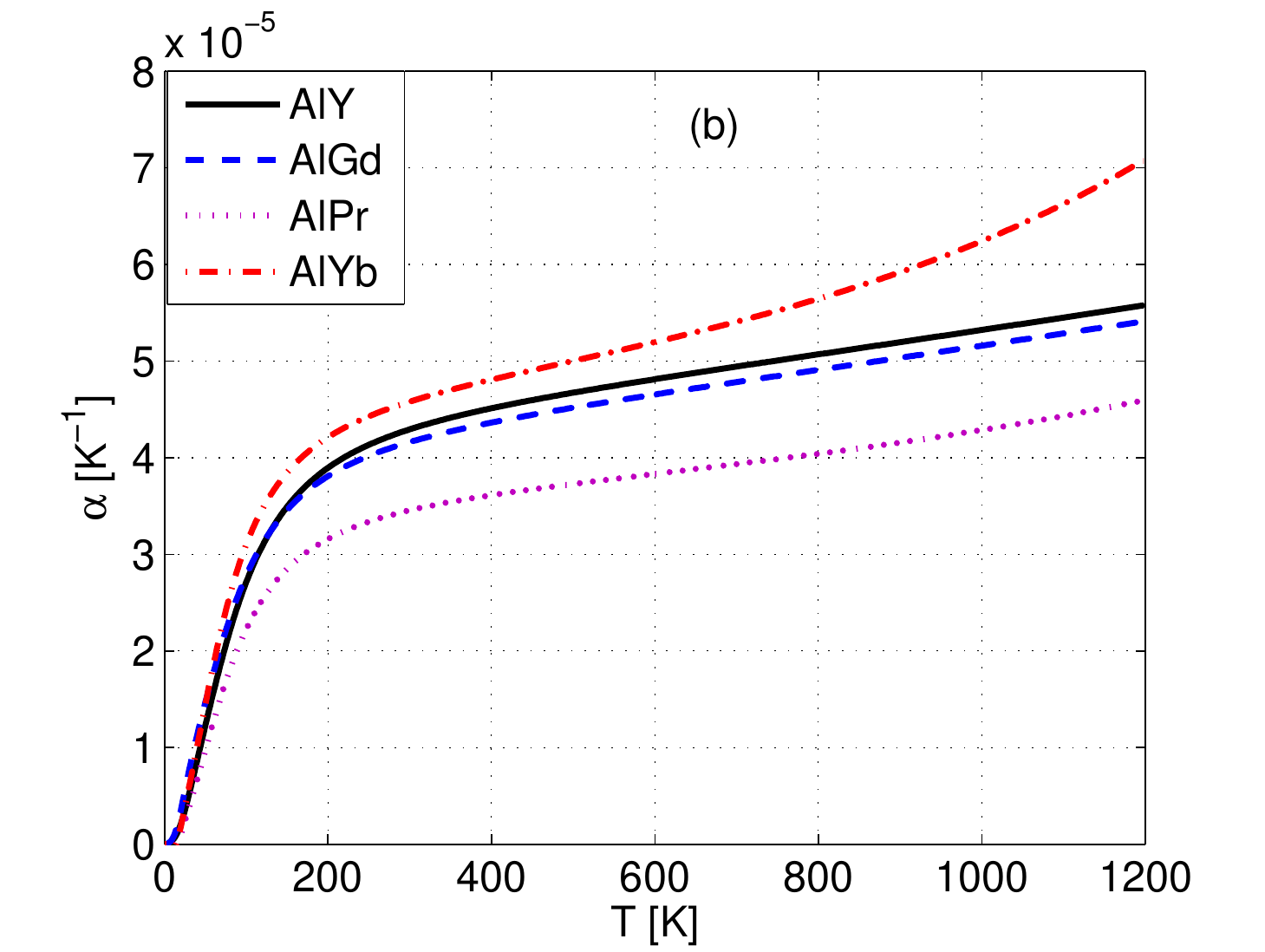}}
\caption{(Color online) (a) Temperature dependence of the linear thermal expansion $\epsilon(T)$ for AlRE(RE=Y, Gd, Pr, and Yb); (b) The coefficients of volume thermal expansion $\alpha$ for AlRE(RE=Y, Gd, Pr, and Yb) as a function of temperature.}
\label{thermal}
\end{figure}

As shown in Figure \ref{EOS}, the equilibrium volume at any temperature corresponds the minimum values of the fitted thermodynamic functions. The thermal expansion is observed as an increase in the equilibrium volume. The linear thermal expansion $\epsilon$ defined by Eq. (\ref{epsilon}) as a function of the four AlRE intermetallics are shown in Figure \ref{thermal}(a). The linear expansions of the four compounds are found AlYb$>$AlY$>$AlGd$>$AlPr, and those of AlY and AlGd are nearly equivalent.  The coefficients of the volume thermal expansion $\alpha$ as a function of temperature are shown in Figure \ref{thermal}(b). With increasing temperature, the thermal expansion grows rapidly up to $\sim200K$, and the slops become smaller and nearly constant at high temperatures except AlYb. In addition, $\alpha$ can be used to estimate the anharmonic effects according to Gr\"{u}neisen relation $\alpha=\gamma C_{V}/VB_{T}$, and Gr\"{u}neisen parameter $\gamma=-d \ln \omega /d \ln V$. In QHA, the phonon frequencies at given lattice parameters are independent of temperature. In real crystal, it is not the case. The accuracy of QHA applied to AlRE (RE=Y, Gd, Pr, Yb) can be verified by experiment in the future.

\subsection {Specific heat and Entropy}
\begin{figure}

\scalebox{0.5}[0.5]{\includegraphics{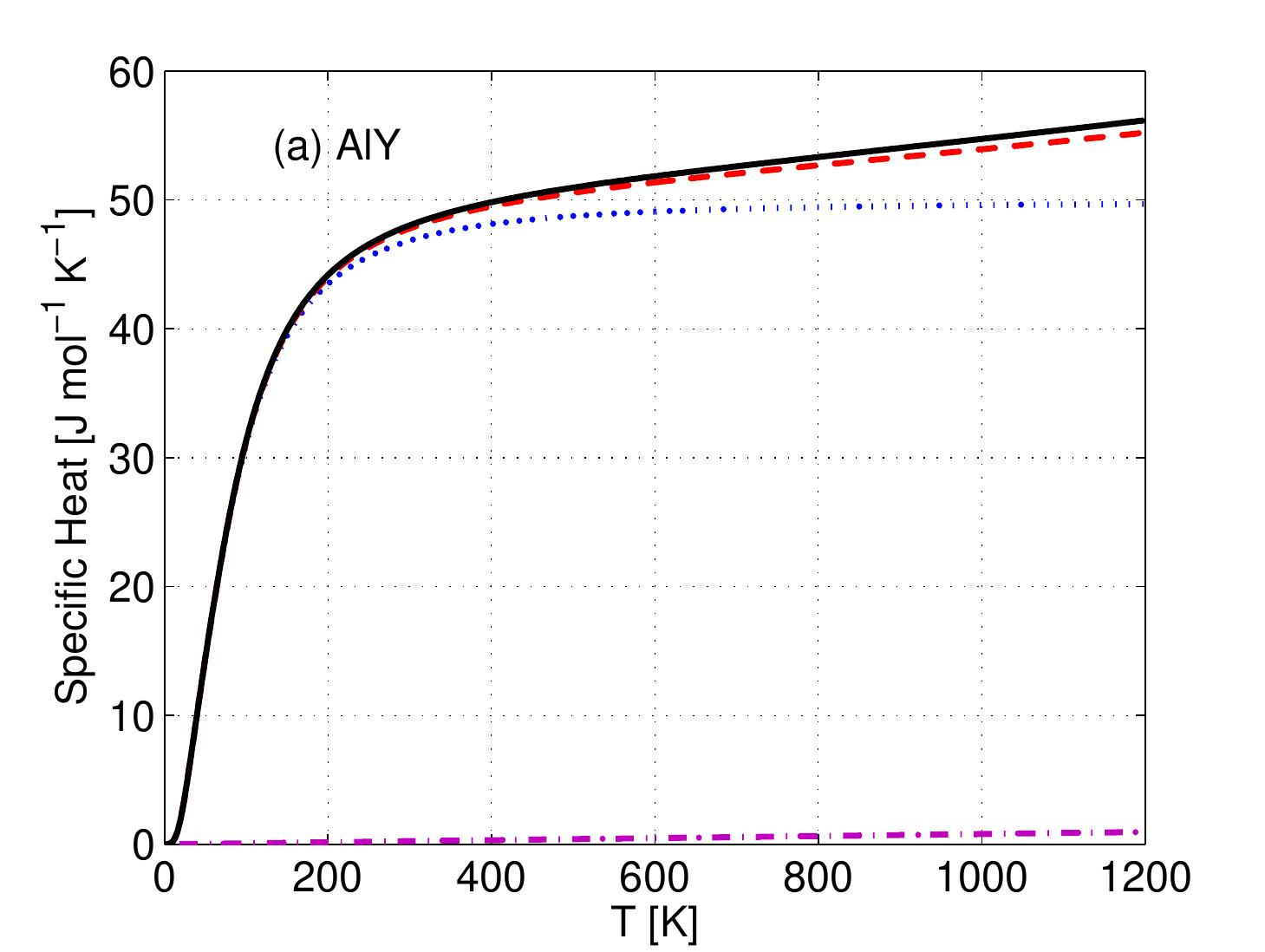}}
\scalebox{0.5}[0.5]{\includegraphics{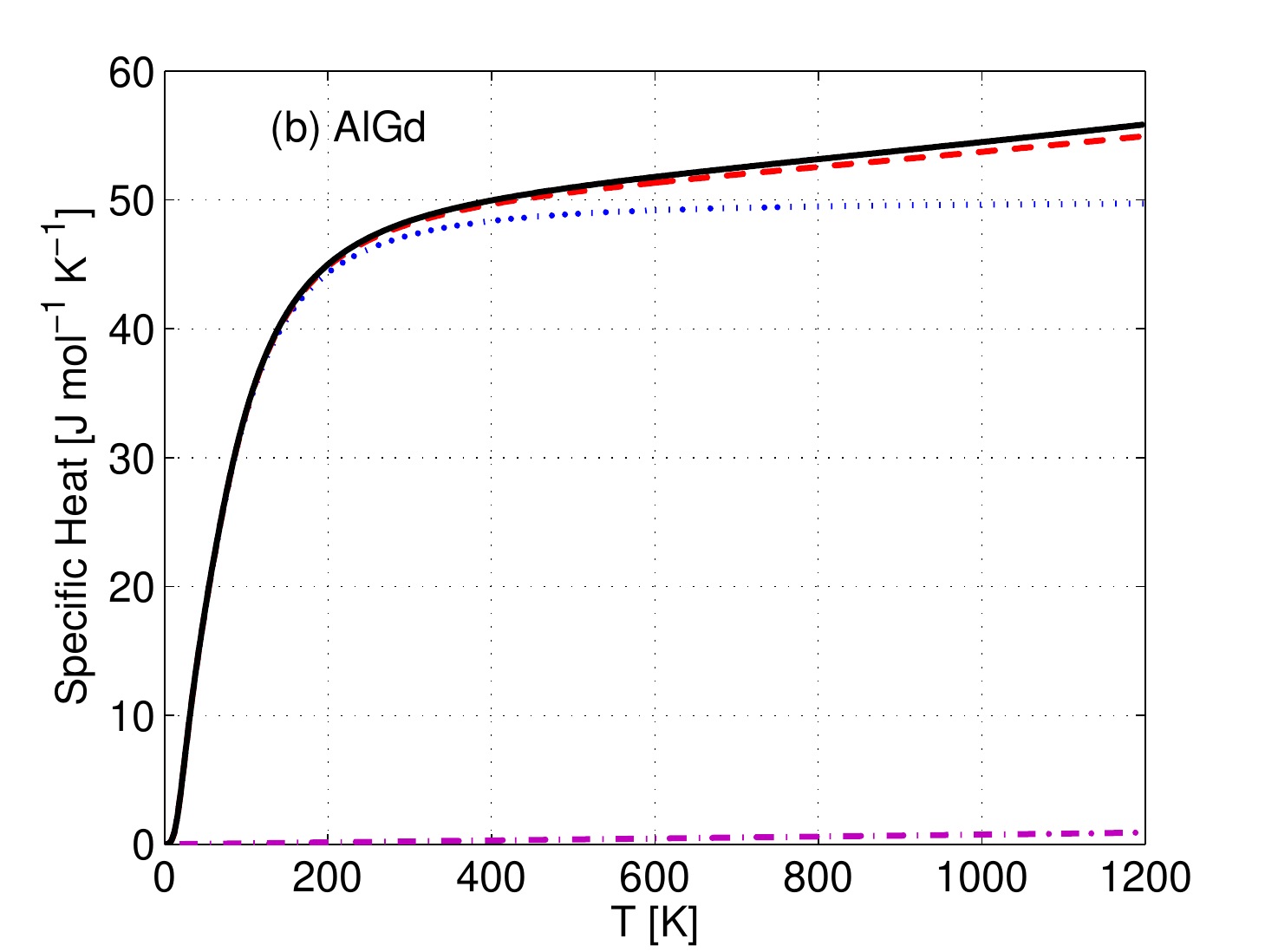}}

\scalebox{0.5}[0.5]{\includegraphics{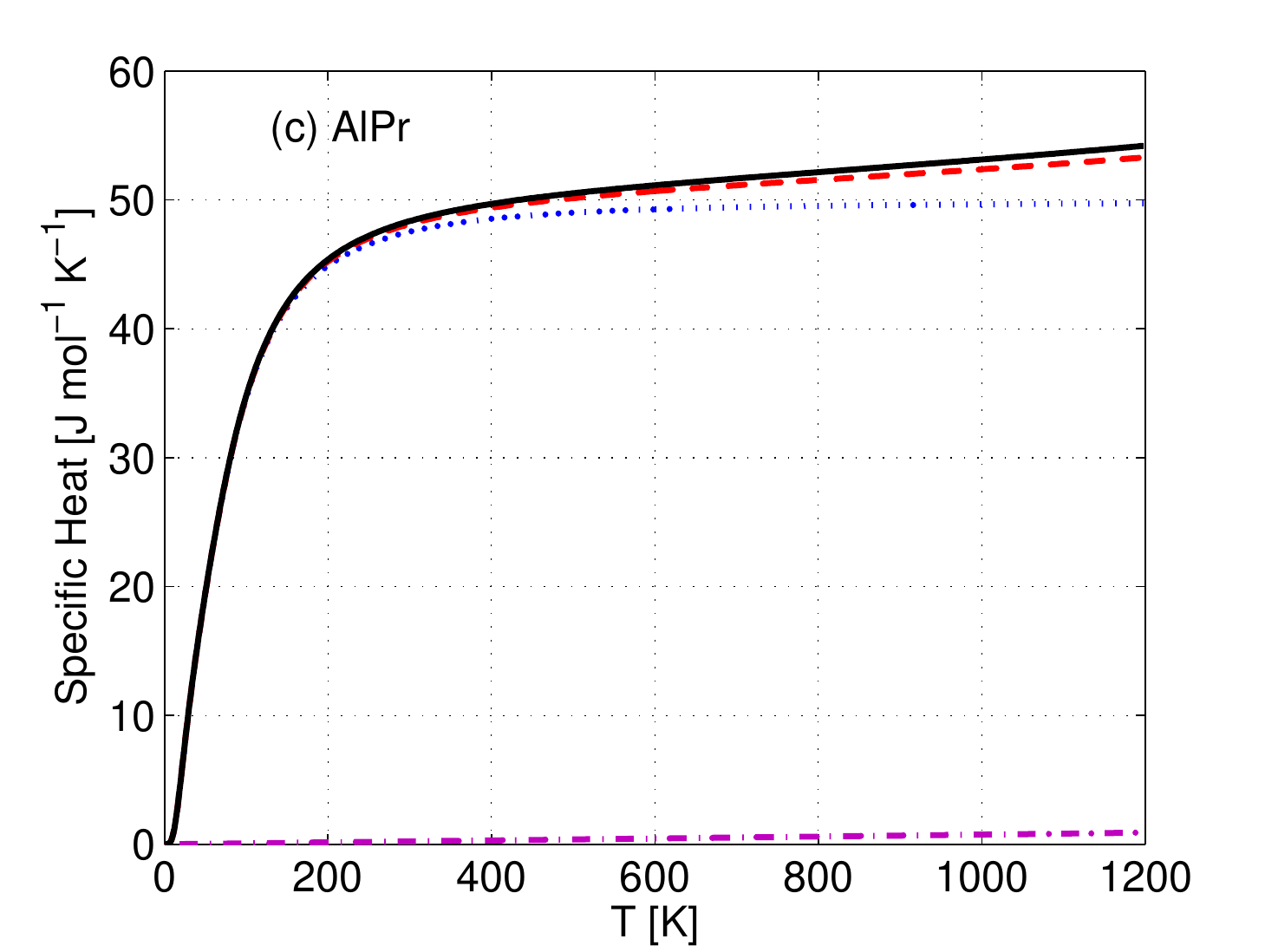}}
\scalebox{0.5}[0.5]{\includegraphics{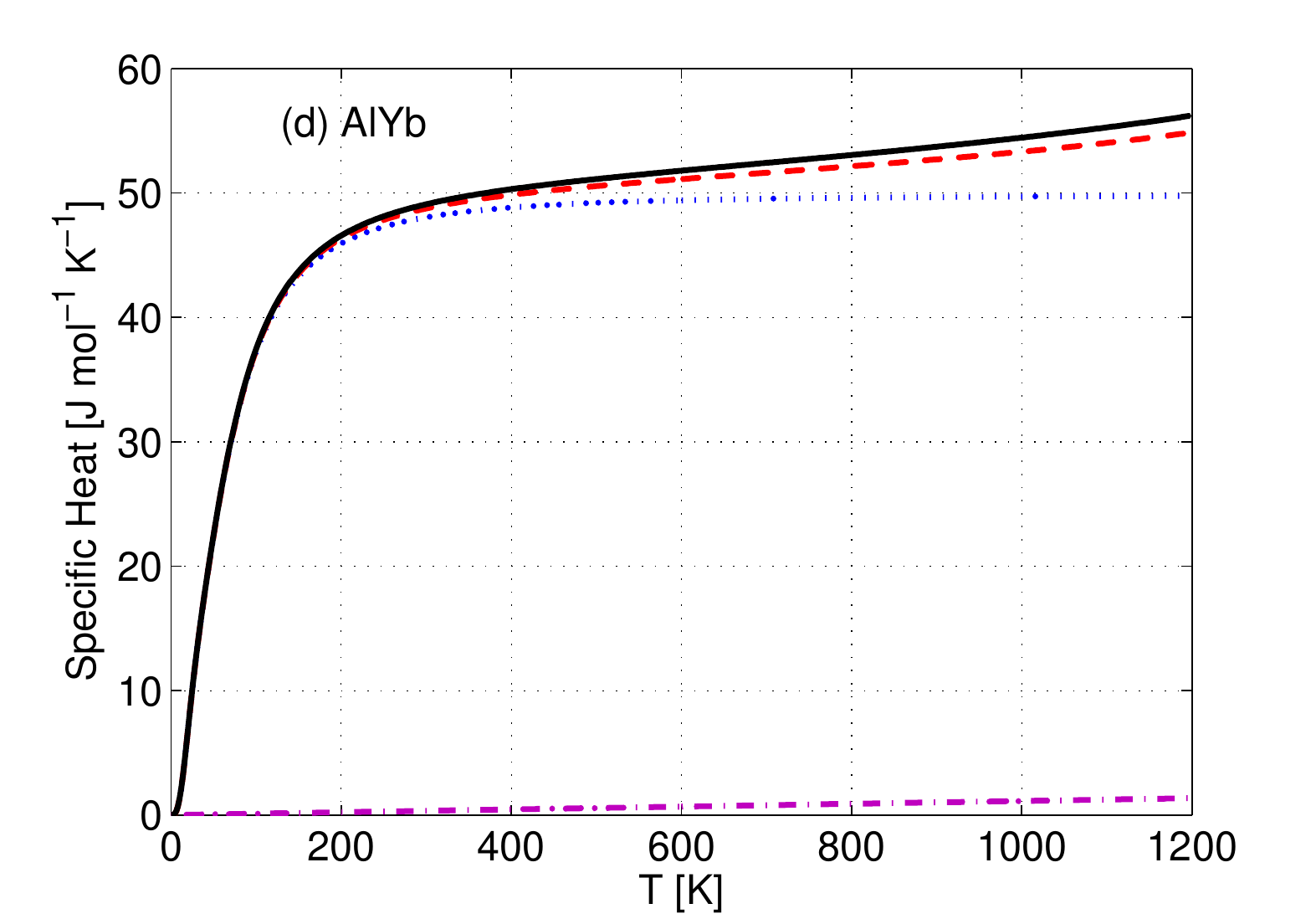}}
\caption{Temperature dependence of heat capacity of (a) AlY, (b) AlGd, (c) AlPr, and (d) AlYb. Solid and dashed curves denote the calculated $C_{p}$, including electronic contribution and not, respectively. Doted and dot-dashed curves show vibrational and electronic $C_{V}$, respectively.}
\label{C}
\end{figure}

\begin{figure}
\scalebox{0.7}[0.7]{\includegraphics{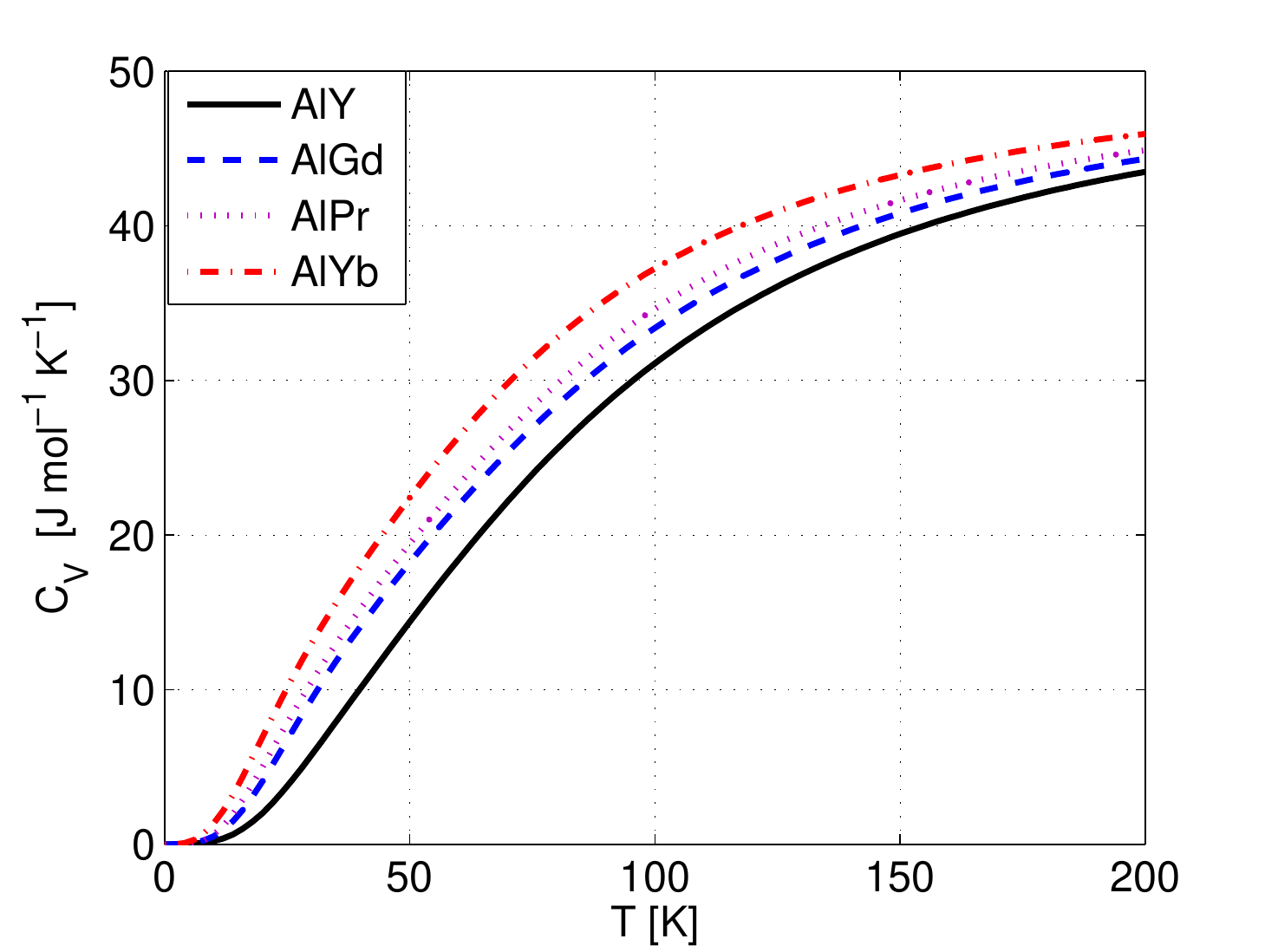}}
\caption{(Color online) The specific heat at constant volume $C_{V}$ for AlRE(RE=Y, Gd, Pr, and Yb) as a function of temperature below 200K. }
\label{Clow}
\end{figure}

Once the phonon spectrum over the entire Brillouin zone is available, the vibrational heat capacity at constant volume $C_{V}^{\mathrm{vib}}$ can be calculated by using Eq. (\ref{cvib}), while the electronic contribution to heat capacity at constant volume $C_{V}^{\mathrm{el}}$ can be obtained from the electronic DOS by using Eq. (\ref{cel}). Then, the specific heat at constant pressure $C_{\mathrm{p}}$ can be computed by Eq. (\ref{cp}). As a comparison, both $C_{p}$ values, including electronic contribution and not, are plotted. We display these results in Figures \ref{C}. At high temperature, $C_{V}^{\mathrm{vib}}$ tends to the classical constant value $6R$ ($R$ is molar gas constant), and $C_{V}^{\mathrm{el}}$ and $C_{\mathrm{p}}$ still increase. Considering thermal electronic contributions to specific heat, we find that $C_{V}^{\mathrm{el}}$ lets $C_{\mathrm{p}}$ suffer a little shift and can be negligible. This character can be understood from the electronic DOS at the Fermi level $n(\varepsilon_{F})$. AlRE intermetallics have low electronic DOS near the Fermi level and the Fermi levels occur at a valley in the curves of electronic DOS \cite{Tao1}. Obviously, $C_{V}^{\mathrm{el}}$ is much smaller than the value $C_{\mathrm{p}}-C_{\mathrm{V}}=\alpha^{2}BVT$. Hence, the electronic excitations affecting the thermal properties is inessential. Comparing with our previous study of MgRE intermetallics \cite{Rui2011}, the electronic specific heat can not be negligible since the Fermi  level for MgRE intermetallics occur above a peak in the electronic DOS \cite{Wu} and the electronic excitations affecting the thermal properties is remarkable. Whether we consider the electronic contribution or not, the theoretical $C_{\mathrm{p}}$ keeps positive slopes obviously. At low temperature below $\sim200K$, the discrepancy between $C_{\mathrm{p}}$ and $C_{\mathrm{V}}$ can be neglected, and $C_{V}$ values increase rapidly.  We show temperature dependence of $C_{V}$ for AlRE(RE=Y, Gd, Pr, and Yb) below 200K as in Figure \ref{Clow}. At low temperature, the specific heats $C_{V}$ of four AlRE intermetallics obey the law of $T^3$. The values of $C_{V}$ are found AlY$<$AlGd$<$AlPr$<$AlYb, and this character can be understood from elementary excitation of phonon relating to the mass of unit cell. We have the order of the mass of unit cell $M_{\mathrm{AlY}}<M_{\mathrm{AlGd}}<M_{\mathrm{AlPr}}<M_{\mathrm{AlYb}}$. At low temperature limitation $T\rightarrow 0K$, the specific heat $C_{V}$ is determined from the vibrations of the low frequencies.

\begin{figure}
\scalebox{0.7}[0.7]{\includegraphics{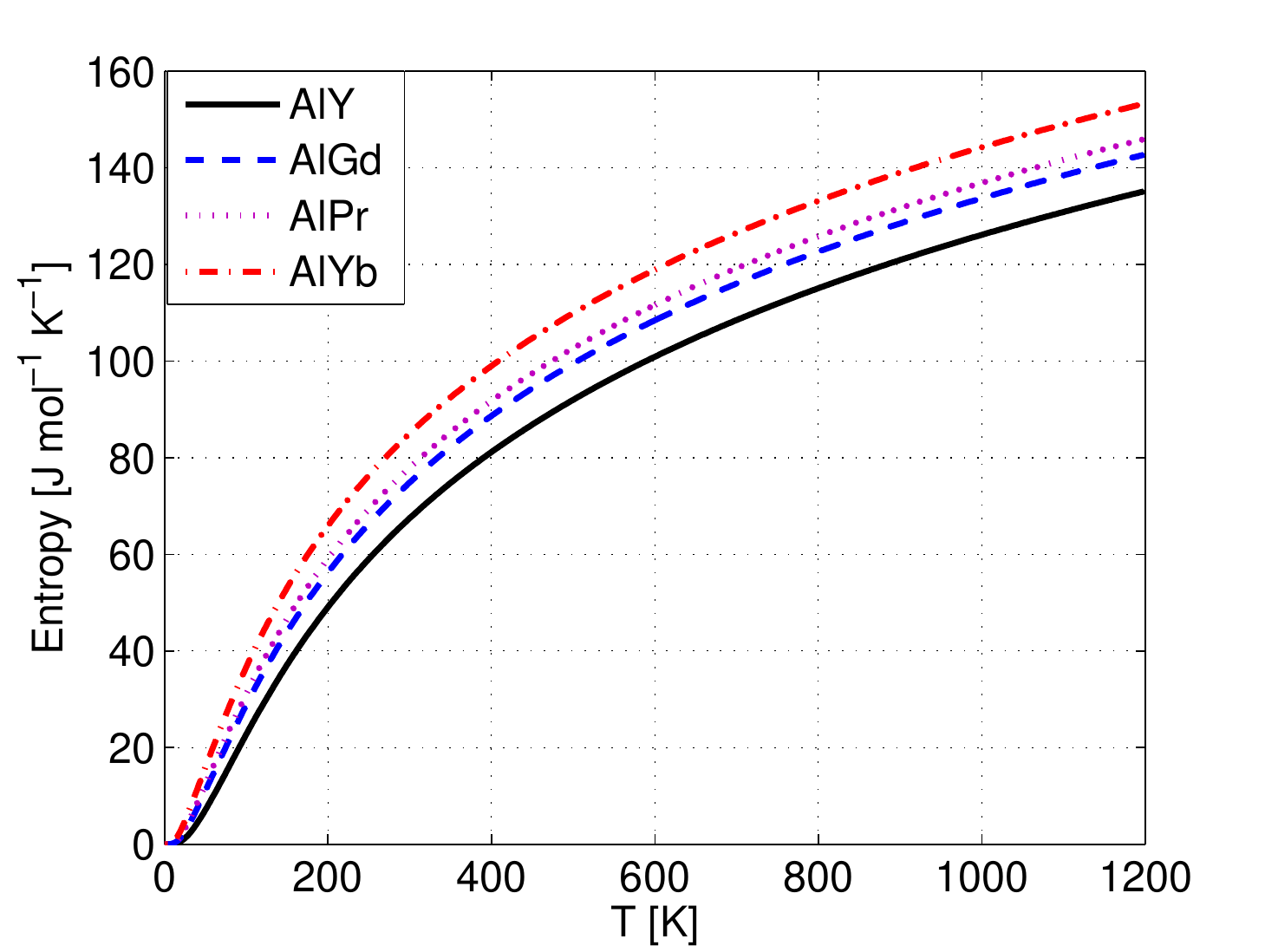}}
\caption{(Color online) The entropy $S$ for AlRE(RE=Y, Gd, Pr, and Yb) as a function of temperature. }
\label{Entropy}
\end{figure}

The entropy as a function of temperature is an important thermodynamic quantity in thermodynamic modeling. The calculation results of entropies for the fore B2-type structures AlRE are shown in Figure \ref{Entropy}. It can be clearly seen that the entropies of the four compounds increase with temperature. Through the temperature range $0-1200K$, the overall observation is that the order of the entropy $S$ AlY$<$AlGd$<$AlPr$<$AlYb. At
the temperature of $300K$, the calculated values of entropy for AlY, AlGd, AlPr, and AlYb are 67.48 $\mathrm{J mol^{-1}K^{-1}}$, 74.87 $\mathrm {J mol^{-1}K^{-1}}$, 77.93 $\mathrm{J mol^{-1}K^{-1}}$, and 85.03 $\mathrm{J mol^{-1}K^{-1}}$.

\section {Conclusions}

In conclusion, the phonon and thermodynamic properties, such as the thermal expansions, the heat capacities at constant volume and constant pressure, the isothermal bulk modulus, and the entropy as function of temperature, of rare-earth-aluminum intermetallics AlRE (RE=Y, Gd, Pr, Yb) with B2-type structure are investigated by DFT and DFPT within the QHA. The phonon spectra and phonon density of states, including the phonon partial density of states and total density of states, have been discussed. Our results demonstrate that the density of states are mostly composed of Al states at the high frequency and composed RE (RE=Y, Gd, Pr, and Yb) states at low frequency. Through the calculated temperature range $0-1200K$, AlY is the most incompressible and AlYb is the most compressible. The differences of the isothermal bulk moduli among the four intermetallics almost remain unchanged with increasing temperature. The thermal expansions of the four compounds are found AlYb$>$AlY$>$AlGd$>$AlPr, and those of AlY and AlGd are nearly equivalent. The specific heats and entropy increase with temperature, while $C_{V}^{\mathrm{vib}}$ tends to a constant and $C_{\mathrm{p}}$ keeps positive slopes at hight temperature. The electronic contribution to the specific heat is discussed, and the presented results show that the thermal electronic excitation affecting the thermal properties is inessential.


 \vskip 2in

\def\refname{{\large\bfseries References}}

\newpage
\begin{table}
\caption{The equilibrium lattice constants $a_{0}$, the isohtermal bulk modulus $B_{0}$, and the pressure derivatives of the isothermal bulk modulus $B_{0}'$ at $T=0K$ for AlRE (RE=Y, Gd, Pr, Yb) in our calculation in comparison with the previous calculated results and the experiment.}

\begin{tabular}{ccccc}
  \hline
   & AlY& AlGd & AlPr & AlYb \\
   \hline
  $a_{0} ({\AA})$ & 3.606, 3.605$^{a}$, 3.759$^{b}$ & 3.635, 3.634$^{a}$, 3.7208$^{c}$& 3.760, 3.759$^{a}$, 3.82$^{d}$ & 3.699, 3.697$^{a}$\\
  $B_{0}$ (GPa) & 62.84, 63.70$^{a}$ & 62.54, 61.72$^{a}$ & 54.18, 55.11$^{a}$ & 38.72, 39.11$^{a}$ \\
  $B_{0}'$ & 3.97, 3.99$^{a}$ & 3.79, 3.81$^{a}$ & 3.86, 3.92$^{a}$ & 4.24, 4.30$^{a}$ \\
  \hline
\end{tabular}

\begin{tabular}{cccccccc}
  \leftline {${}^{a}$Reference\cite{Tao1}}\\
  \leftline {${}^{b}$Reference\cite{Dagerhamn}} \\
  \leftline {${}^{c}$Reference\cite{Baenziger}} \\
  \leftline {${}^{d}$Reference\cite{Kripjakevich}}\\
\end{tabular}
\label{table}
\end{table}

\end{document}